\documentclass[amsmath,amssymb,aps,prl,superscriptaddress,twocolumn,notitlepage,nofootinbib]{revtex4-1}
\usepackage[dvips]{graphicx}
\usepackage[USenglish]{babel}
\usepackage{amsmath,amssymb,amsthm,dsfont,url,braket,hyperref, mathtools,commath}
\usepackage[percent]{overpic}
\usepackage{xcolor}
\usepackage{float}

\newcommand{\ch}[1]{\mathcal{#1}}

\newcommand{\ketbra}[2]{\ket{#1}\!\!\bra{#2}}

\newcommand{\ditv}{\textsc{di-tv}}
\newcommand{\dicc}{\textsc{di-cc}}

\begin{document}

\title{Experimental device-independent tests of quantum
  channels}

\author{Iris Agresti}
\affiliation{Dipartimento di Fisica - Sapienza Universit\`{a} di Roma, P.le Aldo Moro 5, I-00185 Roma, Italy}

  \author{Davide Poderini}
\affiliation{Dipartimento di Fisica - Sapienza Universit\`{a} di Roma, P.le Aldo Moro 5, I-00185 Roma, Italy}
  
  \author{Gonzalo Carvacho}
\affiliation{Dipartimento di Fisica - Sapienza Universit\`{a} di Roma, P.le Aldo Moro 5, I-00185 Roma, Italy}
  
  \author{Leopoldo Sarra}
\affiliation{Dipartimento di Fisica - Sapienza Universit\`{a} di Roma, P.le Aldo Moro 5, I-00185 Roma, Italy}

 \author{Rafael Chaves}
\affiliation{International Institute of Physics, Universidade Federal do Rio Grande do Norte, Campus Universitario, Lagoa Nova, Natal-RN 59078-970, Brazil}

\author{Francesco Buscemi}
\email{buscemi@is.nagoya-u.ac.jp}
\affiliation{Department of Mathematical Informatics, Nagoya University, Chikusa-ku, Nagoya, 464-8601, Japan}

\author{Michele Dall'Arno}
\email{cqtmda@nus.edu.sg}
\affiliation{Centre for  Quantum Technologies, National University of  Singapore, 3 Science Drive  2, 117543, Singapore}

\author{Fabio Sciarrino}
\email{fabio.sciarrino@uniroma1.it}
\affiliation{Dipartimento di Fisica - Sapienza Universit\`{a} di Roma, P.le Aldo Moro 5, I-00185 Roma, Italy}

\date{\today}

\begin{abstract}
Quantum tomography is currently the mainly employed method to assess the information of a system and therefore plays a fundamental role when trying to characterize the action of a particular channel. Nonetheless, quantum tomography requires the trust that the devices used in the laboratory perform state generation and measurements correctly. This work is based on the theoretical framework for the device-independent inference of quantum channels that was recently developed and experimentally implemented with
superconducting qubits in [Dall'Arno, Buscemi, Vedral, arXiv:1805.01159] and [Dall'Arno, Brandsen, Buscemi, PRSA 473, 20160721 (2017)]. Here, we present a complete experimental test on a photonic setup of two device-independent quantum channels falsification and characterization protocols to analyze, validate, and enhance the results obtained by conventional quantum process tomography. This framework has fundamental implications in quantum information processing and may also lead to the development of new methods removing the assumptions typically taken for granted in all the previous protocols. 
\end{abstract}

\maketitle

Measurements are essential to acquire information about physical systems and its dynamics in any experimental science. In quantum physics, in particular, the importance of measurements is promoted even further since they perturb the quantum system under scrutiny and thus require a new understanding of how to connect observed empirical data with the underlying quantum description of nature. To cope with that, we can rely on quantum tomography \cite{tomo1,tomo2,tomo3}, a general procedure to reconstruct quantum states and channels from the statistics obtained by measurements on ensembles of quantum systems. However, how can one guarantee that the measurement apparatus is measuring what it is supposed to? In practice, experimental errors are unavoidable and such deviations from an ideal scenario not only can lead to the reconstruction of unphysical states \cite{unph} but also imply false positives in entanglement detection \cite{ent1, ent2, ent3} and compromise the security in quantum cryptography protocols \cite{cryp1,cryp2, cryp0, cryp00}.

Strikingly, with the emergence of quantum information science, a new paradigm has been established for the processing of information. This is the so-called Device-Independent (\textsc{di}) approach \cite{div, dient, div2,Dal17, DBBV17}, a framework where conclusions and hypotheses about the system of interest can be established without the need of a precise knowledge of the measurement apparatus/devices. The prototypical example of how the \textsc{di} reasoning works is given by Bell's theorem \cite{Bell1,Bell2}, which implies experimentally testable inequalities, whose violation certifies the presence of entanglement and provides further information about the quantum state, such as its dimension \cite{dim1,dim2} or fidelity with a maximally entangled state \cite{div3}. In other words, even with no information whatsoever about what measurements are being performed, general features of the quantum state can be recovered. A natural question is then whether the \textsc{di} approach can also be adopted within the other pillar of quantum tomography: the reconstruction of quantum channels.

\begin{figure}[hbt]
    \includegraphics[width=.5\textwidth]{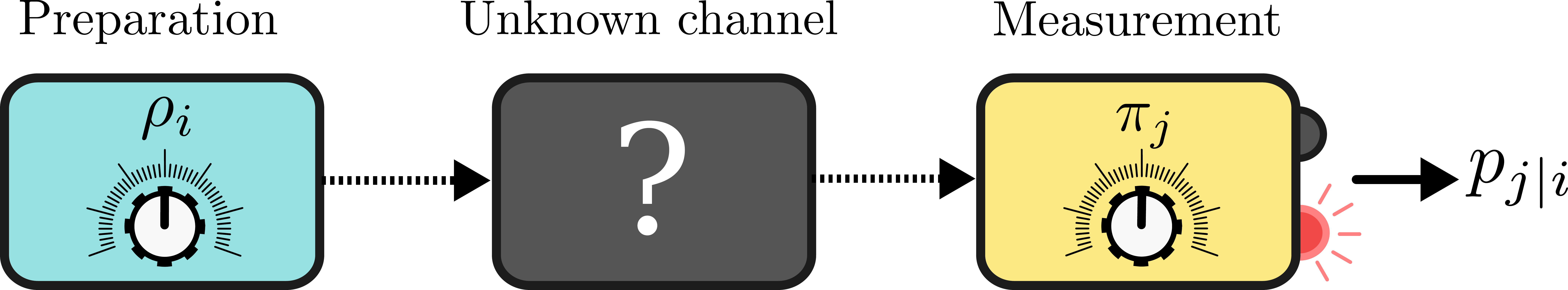}
    \caption{\textbf{Schematic representation of a measurement process.} 
        A state $\rho_i$ is prepared and sent through the
        unknown channel, after which we perform a measurement $\pi_j$.
        Repeating the process for different combinations of states and
        measurements we get the correlations $p_{j|i}$.}
 \label{fig:measurement}
\end{figure}

\begin{figure}[hbt!]
    \includegraphics[width=.47\textwidth]{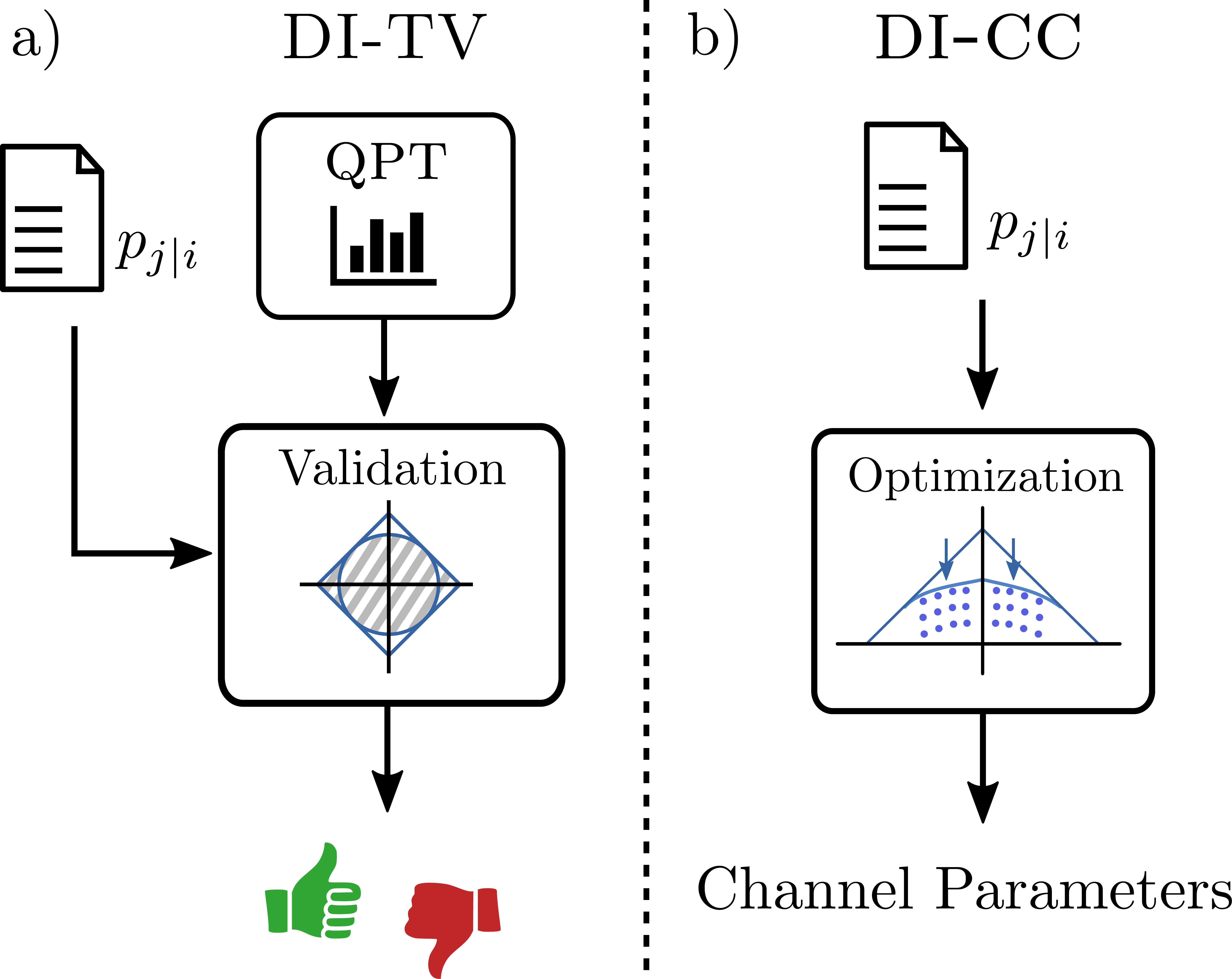}
    \caption{\textbf{Summary of the DI tests.}
    \ditv{} (device-independent tomography validation): We
    first perform a standard quantum process tomography (QPT) on the
    channel, using maximum likelihood estimation \cite{likel,hrad}. We then find the set of
    correlations compatible with the reconstructed channel and perform additional measurements, obtaining correlations $p_{j|i}$ that can invalidate the QPT if they do not belong to the expected set.
    b) \dicc{} (device-independent channel characterization): In this
    case we do not perform the QPT, instead we collect a fixed number of
    correlations $p_{j|i}$, spanning uniformly the Bloch sphere with the states
   $\rho_i$ and measurements $\pi_j$ (but, as opposed to QPT, without any assumption on their actual physical implementation). Then we find the channel parameters (see ~\cite{dell2018,ditest} and Supplementary Material) that best fit the set of experimentally found correlations.}
 \label{fig:analisys}
\end{figure}

This question was very recently answered on the affirmative by Dall'Arno et al.
in Refs.~\cite{dell2018,ditest}, where a theoretical framework for the
device-independent inference of unknown quantum channels, given as a black-box,
was derived and experimentally implemented with superconducting qubits.
In this article, we adopt the theoretical tools developed in \cite{dell2018, ditest} and implement a \textsc{di} validation test of a quantum process tomography, i.e. device-independent tomography validation (\ditv{}). This protocol is schematically represented in Fig.\ref{fig:analisys}a and can be adopted after a usual tomography has been performed on a given quantum channel, in order to falsify or validate the channel reconstruction obtained from the data. As a further step, we propose an algorithm to provide a confidence range within which the tomography is validated (see also \cite{dell2018} for the discussion of a different algorithm for the same purpose). Following the \ditv{}, we implement a second protocol, depicted in Fig.\ref{fig:analisys}b: a device-independent channel characterization (\dicc{}), where we have no information about the nature of the channel and, by preparing random quantum states and measuring random observables, we can determine the quantum channel equivalence class that is the most compatible with the measurement data. 
The two protocols, recently introduced in \cite{dell2018}
are applied here on two different types of quantum channels, the amplitude damping and the dephased amplitude damping channels, by exploiting a photonic setup and using a qubit encoded in the polarization degree of freedom. 

The scenario of interest is depicted in Fig.~\ref{fig:measurement}. In order to characterize an unknown quantum channel $\ch{X}$ via tomography, we prepare a number of known quantum states $\rho_i$ that evolve as $\ch{X}(\rho_i)$ and that are then measured according to a measurement operator $\pi_j$ associated with outcome $j$.
For example, for a single qubit, quantum process tomography (QPT) \cite{qpt1} requires the preparation of the states $\ket{0}$, $\ket{1}$, $\ket{+} = (\ket{0} + \ket{1})/\sqrt{2}$, $\ket{R} = (\ket{0} - i\ket{1})/\sqrt{2}$ and the projection onto the eigenstates of the Pauli matrices $\sigma_{x}$, $\sigma_{y}$ and $\sigma_{z}$. By the Born's rule we associate the observed statistics with the state and measurements via $p_{j|i} = \mathrm{Tr}\left[\pi_j \ch{X}(\rho_i) \right]$ and, if we know $\rho_i$ and $\pi_j$, this equation can be inverted to find the expression for the channel $\ch{X}$. 
However, if we have no information (or trust) about the states and measurements, what kind of relevant information can be extracted from the observed statistics $p_{j|i}$? 
As proposed in \cite{ditest, dell2018}, a given quantum channel defines a convex set of correlations $p_{j|i}$ that are compatible with it. Thus, if the experiment produces a point $p_{j|i}$ outside the boundary of the correlation set of a channel, taken as hypothesis (for example arising from the quantum process tomography), we can unambiguously exclude this channel as potential candidate for our reconstruction. 
Alternatively, if the measured correlation points $p_{j|i}$ fall within the set, the hypothesis is not falsified and one can move further, searching for a confidence level within which the quantum process tomography prediction is validated. Moreover, overlooking any device-dependent channel reconstruction, one can find the best quantum channel fitting the data, as the ``minimal surviving hypothesis'', that is, the channel $\ch{X}^*$ whose boundary contains all the observed correlations set, but minimizing the surrounded volume \cite{dell2018}. Note that this protocol cannot unambiguously determine all of the characterizing parameters of the quantum channel, so it actually enables to single out the equivalence class to which it belongs (see Supplementary Material of \cite{dell2018}).
Considering the case where $i,j=0,1$, that is two possible state preparations and a dichotomic measurement, analytical expressions characterizing a large class of qubit channels, which are invariant under the dihedral group $D_2$, have been obtained \cite{ditest,dell2018}. This class of channels can be represented by four real numbers $\left(d_1, d_2, d_3, c_3 \right)$  and maps the Bloch sphere into an ellipsoid translated along one of its own axis. 
This transformation is pictorially represented in Fig.~\ref{fig:sphere} (see Supplementary Material for further details), which shows that geometrically $d_{1}$, $d_{2}$ and $d_{3}$ represent the ellipsoid's axes, and $c_3$ the translation along the $z$ axis.
\begin{figure}[h!]
  \includegraphics[width=.3\textwidth]{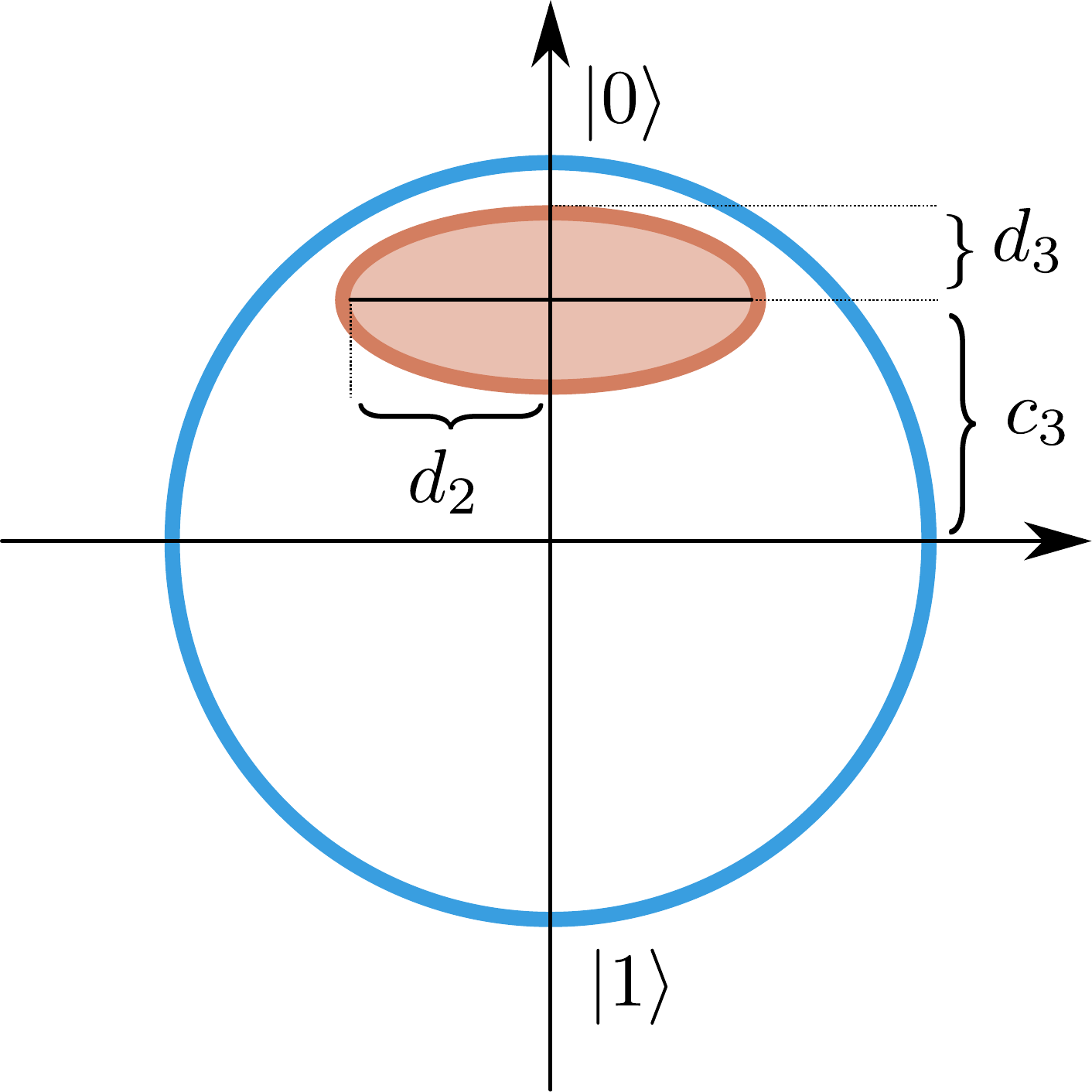}
  \caption{{\bf Bloch-sphere representation} of
    the action of  a general qubit dihedrally-covariant
    channel.   The  Bloch-sphere   is  mapped  into  an
    ellipsoid with  semi-axis $d_1$, $d_2$,  and $d_3$,
    which  is  translated along  one  of  its own  axis
    ($d_3$, in  the figure) by a  distance $c_3$.}
\label{fig:sphere}
\end{figure}

\begin{figure*}[t]
 \includegraphics[width=\textwidth]{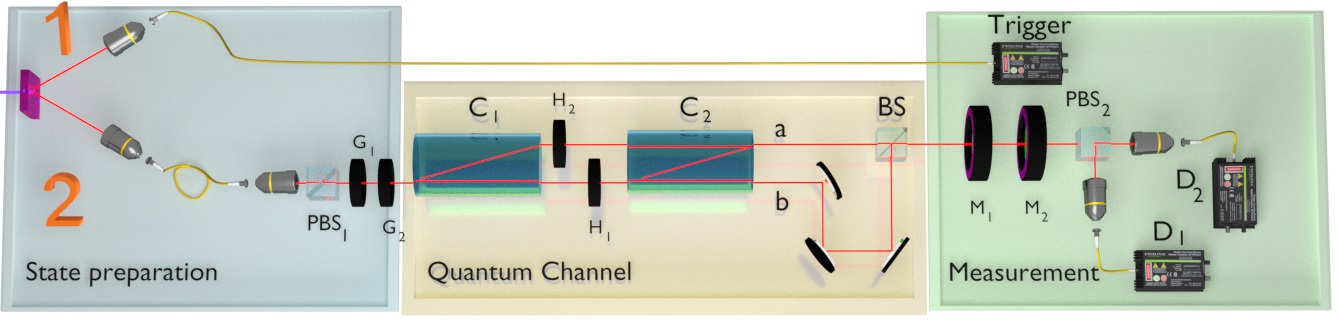}
 \caption{\textbf{Experimental apparatus.} In the preparation stage a couple of photons is generated by a spontaneous parametric down-conversion (SPDC) process in a non-linear crystal. Photon 1 will act as a trigger while photon 2's polarization encodes the particular qubit to be analyzed. The state preparation is performed by a polarizing beam splitter (PBS$_{1}$), a quarter-wave plate ($G_{1}$) and a half-wave plate ($G_{2}$), which rotates the  horizontal polarization of incoming photons in order to generate
 $\sqrt[]{\omega}\ket{0}$+$\sqrt[]{1-\omega}\ket{1}$. After the preparation of the state, photon 2 is sent to the amplitude damping channel section (yellow stage) composed by a first calcite crystal ($C_{1}$), two half-wave plates ($H_{1}$ and $H_{2}$) one for each generated path and another calcite crystal ($C_{2}$) to recombine the paths. Setting the angle of $H_{1}$ 
  at $45^o$ we are able to close the calcite interferometer thus sending the photons to the output ``a".  Conversely, when $H_{1}$ 
 is at $ 0^o$, photon 2 comes out in path ``b" projected onto the vertical polarization $\ket{V}$. The two paths ``a" and ``b" are incoherently recombined at the beam splitter (BS). Intermediate angles between $0^o$ and $45^o$ for $H_{2}$ 
 and $H_{1}$ 
 allow to tune the contribution $\ket{H}$ and $\ket{V}$ to be recombined in
 the BS. The measurement stage is composed by a quarter-wave plate ($M_{1}$), a half-wave plate ($M_{2}$) and a PBS, after which the photons are coupled into single-mode (SM) fibers connected to single photon detectors.}
 \label{fig:calcite}
\end{figure*}

\textbf{\emph{Device-independent tomography validation--}} Now, let us consider the goal to characterize an implemented quantum channel through a tomography, but not trusting our apparatus and therefore requiring a device-independent validation to state its correct functioning. Sticking to the hypothesis of $D_2$ covariance, we perform the QPT, restricting to the hypothesis of dihedral covariance (see Supplementary Material), and find the four parameters ($d_{1}$, $d_{2}$, $d_{3}$, $c_{3}$) that identify our channel. This restriction is fully supported by our experimental evidence, indeed the fidelity between the general and the $D_{2}$ covariance restricted quantum process tomographies, performed on all the implemented channels, has been found to be always higher than 0.99. As following step, we are able to reconstruct the hypothetical channel's boundary of the set of input/output correlations \cite{ditest,dell2018}. The tomographic experiment is device-independently proved to be trustworthy if, uniformly spanning the whole set of correlations, all the observed data fall within the boundary. Otherwise, the quantum process tomography is falsified. In principle, it is possible to exploit the same data set, both for the quantum process tomography and the DI validation. Indeed, in the latter case, the data would be interpreted as bare correlations, without any assumptions about states and measurements. Clearly, however, a better validation includes additional experimental data probing the boundary of the set defined by the channel being validated. After a tomographic reconstruction is validated, we need to quantify the quality of our tomographic reconstruction, since, if the boundary lies too far from the observed points, the hypothesis, although in principle validated, would not be really supported by the data. 
In other words, a good hypothesis is an ``almost  falsified one'': its boundaries should enclose all the observed correlations, but not much more \cite{dell2018}.

\begin{figure*}[t!]
\centering
 \includegraphics[width=.9\textwidth]{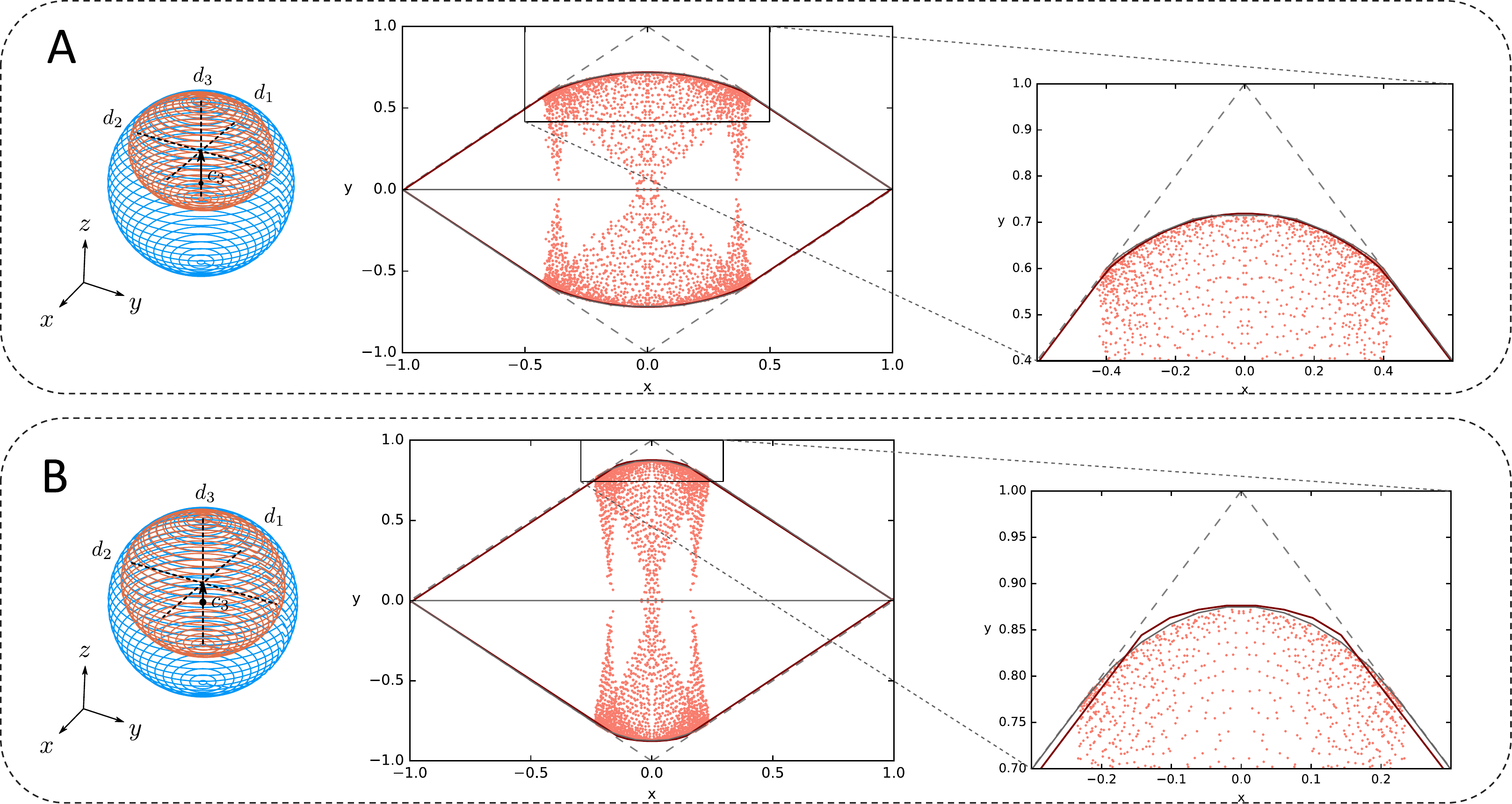}
 \caption{\textbf{Device Independent Test of a Quantum Process Tomography.}  From a QPT on two channels, imposing $D_2$ covariance, we evaluate the four characteristic parameters ($d_{1}$, $d_{2}$, $d_{3}$ and $c_{3}$). The action of the channels on the Bloch sphere is shown on the left, where the orange ellipsoid represents the deformation of the original unit sphere, in blue. The correlations set is surrounded by the red curve. To certificate the QPT, we feed our channel with orthogonal pure states belonging to the XZ plane of the Bloch sphere. In order to cover a significant part of the correlations set, we test all possible combinations of states and measurements. The blank corners within the curves, which are trivial for our analysis (can be obtained by a convex combination of the other points), are not covered, since the channel was fed only with orthogonal states and only orthogonal projection measurements were performed.
 }
\label{qptv}
\end{figure*}

\begin{table*}[t]
\tabcolsep=3mm
\begin{tabular}{ccccc}
\toprule
 & ($d_{1}$, $d_{2}$, $d_{3}$, $c_{3}$) & Parameters' variation range & $\Delta$ & $\mu$\\
\colrule
\textbf{A} &  (0.719, 0.791, 0.596, 0.397) & (-, -), (-0.025, +0.051), (-0.038, +0.023), (-0.028, +0.034) & 0.004 & 0.692\\
\textbf{B} &  (0.815, 0.877, 0.791, 0.231)   & (-, -), (-0.010, +0.020), (-0.025, +0.014), (-0.012, +0.029) & 0.005 & 0.763\\
\botrule
\end{tabular}
\caption{\textbf{Device-Independent Tomography Validation.} Characteristic parameters  ($d_{1}$, $d_{2}$, $d_{3}$, $c_{3}$) obtained for the two channels \textbf{A} and \textbf{B}. The confidence level indicators given by tomography validation are in the right two columns: the parameters' variation range within which the QPT would still be validated, and the $\Delta$, which quantifies the relative difference between the two area of the correlations sets. In the last column on the right we report the $\mu$ parameters, which define the equivalence class to which each of the two channels belongs.}
\label{paramtable}
\end{table*}

This protocol was applied to experimental data, in order to validate quantum process tomography on 1-qubit quantum channels. The implementation, as well as the characterization of the channels, was performed exploiting the photonic platform in Fig.\ref{fig:calcite}. 
The setup can be seen as made of three parts: (i) photonic state preparation, (ii) quantum channel, (iii) measurement station. In part (i), the desired state is prepared encoding one qubit in the polarization degree of freedom of a photon which goes through an apparatus made of a polarizing beam splitter ($PBS_{1}$), followed by a quarter- and half- wave plate ($G_{1}$ and $G_{2}$). The photon is generated by a heralded photon source making use of a spontaneous parametric down-conversion (SPDC) process, where the second photon of the pair is used as a trigger. In part (ii), our quantum channel is made of two birefringent calcite crystals ($C_{1}$ and $C_{2}$), followed by a beam splitter, and has the aim of introducing a desired amount $\lambda$ of decoherence between path a and b, depending on the rotation of the half-wave plate $H_{2}$. The measurement station in part (iii) allows to perform projective measurements, through the sequence of a quarter- and a half- wave plate ($M_{1}$ and $M_{2}$), followed by a $PBS_{2}$ (see Supplementary Material).

By exploiting the aforementioned apparatus, we implemented an amplitude damping channel $\ch{X}^{\mathrm{AD}}$ acting as
\begin{align*}
  \mathcal{A}_\lambda(\rho)  := A_0  \rho A_0^\dagger  + A_1
  \rho A_1^\dagger, \quad \begin{cases} A_0 = \ketbra{0}{0}
    + \sqrt{1-\lambda}\ketbra{1}{1},\\  A_1 = \sqrt{\lambda}
    \ketbra{0}{1}, \end{cases}
\end{align*}
with varying efficiency (i.e. the amount of noise) of the channel as quantified by a parameter $\lambda$.
First, we performed a QPT over this channel, whose estimated action on the Bloch sphere is depicted in orange, in Fig.~\ref{qptv}a and Fig.~\ref{qptv}b. The noise corresponds to deformations of the Bloch sphere: the size of the orange ellipsoid is inversely proportional to the noise strength $\lambda$. Through the QPT, we evaluated the four characteristic parameters ($d_{1}, d_{2}, d_{3}, c_{3}$), reported in Tab.\ref{paramtable}. Using the results in \cite{ditest,dell2018} we could then compute the boundary of the set of input/output correlations and the states and measurements allowing us to probe this boundary (see Supplementary Material). To validate the QPT hypothesis, we prepared 29 different pairs of orthogonal states and projected each of these couples onto 29 pairs of orthogonal directions, in order to span a significant part of correlations set, 
using in total 841 combinations of states and measurements for each plot. 

Both tomographies were validated, since, within 2$\sigma$, all the points lie inside the boundary, as it is shown in the plots in Fig.\ref{qptv}a,b. The confidence level was estimated in two ways. First, as proposed in \cite{dell2018}, by evaluating the relative difference between the union and the intersection of the two correlation sets ($\Delta$). The obtained values were respectively 0.004 (A) and 0.005 (B).  
A second evaluated quantity was the variation range associated to each of the parameters ($d_{2}$, $d_{3}$, $c_{3}$) for each channel, reported in Tab.\ref{paramtable}. This is the range of values within which the characterizing parameters would still allow to validate the QPT. This kind of uncertainty was achieved imposing the two following conditions: for every parameter set in the given range, the QPT boundary strictly surrounds all the experimental points (within 1$\sigma$), and the $\Delta$ parameter cannot be larger than 2 times its original value.
As mentioned before, our device-independent procedure cannot determine all of the four parameters ($d_{1}$, $d_{2}$, $d_{3}$, $c_{3}$), specifically, it is insensitive to either $d_{1}$ or $d_{2}$ (see \cite{dell2018}). We choose $d_{1}$ as the parameter that the procedure is insensitive to, so its uncertainty is not reported. Adopting this protocol, with no assumptions on the implementation nor on the state generation/measurement execution, we were able to recognize the equivalence class of our channel in a fully device-independent way. Indeed, the device-independent boundary reconstruction can be the same for different quantum channels, as reported in \cite{dell2018}.
We can, therefore, define the following quantity:
  $\mu  \left( \ch{C}_{d_{1}, d_{2}, d_{3},\vec{c}}  \right)  =  (1-c_3)(d_2^2-d_3^2)/(c_3 d_3^2)$;
this parameter specifies the equivalence class to which the reconstructed quantum channel belongs. In our case, for both the implemented quantum channels, $0 \leq \mu \leq 1$, as reported in Tab.\ref{paramtable}. Within this regime, two device independent reconstructions $\mathcal{C}_{(d_{1}, d_{2}, d_{3}, \vec{c})}$ and $\mathcal{C}_{(d'_{1}, d'_{2}, d'_{3}, \vec{c}')}$ are indistinguishable when $(d_2, d_3, c_3)= (d_2', d_3', c_3')$. This allows to recognize whether our apparatus is working correctly, generating the correct states and performing the required measurements. As pointed out above, the most plausible hypothesis describing a set of correlations is the one whose boundary encloses all the correlation points while being as close as possible to them.

\textbf{\emph{Device-independent channel characterization}} The second implemented protocol, the device-independent Channel Characterization (\dicc{}), naturally arises if the above insight is lifted to the level of a guiding principle to choose the most plausible hypothesis compatible with the observed data, as discussed in Ref.~\cite{dell2018}. The idea here is to single out the ``minimal
 surviving hypothesis'', that is the channel $\ch{X}^*$ whose boundary encloses the smallest volume $V(\ch{X}^*)$ in correlation space but still contains all the observed correlations set $\mathcal{D} = \{(x_k, y_k )\}$.
 According to this idea, first introduced in \cite{dell2018}, the best candidate channel is given by
 \begin{align}
   \ch{X}^* =  \arg\min_{\substack{\ch{X} \ : \ \mathbb{D}_2\\\mathcal{D} \subseteq
       \mathcal{S}(\ch{X})}} V(\ch{X}).
       \label{minimize}
 \end{align}
 where $\mathcal{S}(\ch{X})$ is the set of correlations associated with the channel $\ch{X}$.
 This set can be spanned sending states and performing projections which are uniformly distributed over the Bloch sphere. 
 
\begin{figure}[h!]
\includegraphics[width=0.47\textwidth]{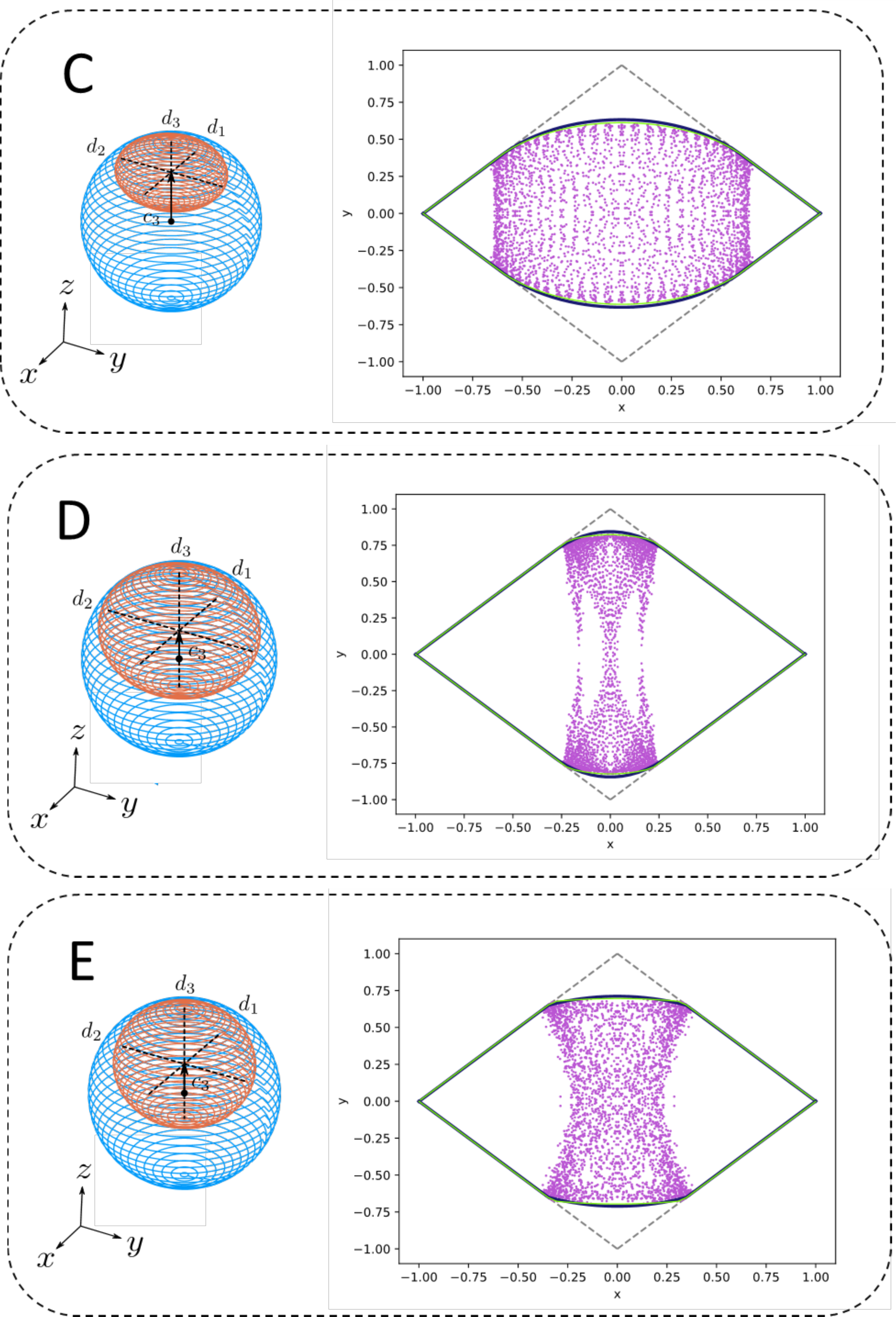}
  \caption{\textbf{Device Independent Channel Characterization of input-output correlations (DI-CC).} Characterization of three channels, implemented with different $\lambda$ parameters. The purple dots are obtained feeding orthogonal pure states belonging to the XZ plane of the Bloch sphere to our channel. The green curves surround the correlations of the \dicc{} while the blue ones are those obtained by the QPT, imposing $D_2$ covariance. For each plot, the action on the Bloch sphere given by the QPT is shown in orange on the left. In the plot the empty sectors can be trivially obtained using non orthogonal states and projectors and are not necessary for our analysis.}
  \label{dicc}
 \end{figure}

 In the experiment, we covered a significant part of this set, through 841 pairs of state/measurement combinations, as in the previous case.
 Here, the aim is to characterize the implemented channel without relying on any device-dependent procedure.
 After the parametrization of the experimental data, the boundary of the correlation set is obtained through a minimization algorithm, based on sequential quadratic programming \cite{slsqp}, with the aim of finding the minimum area which contains all the correlation set, compatible with the constraints imposed by the dihedral covariance of the channel (see the Supplementary Material).
 Exploiting this protocol we can identify, in an entirely device-independent way, the equivalence class to which our channel belongs (see \cite{dell2018}), through three of the four characterizing parameters ($d_{2}$, $d_{3}$, $c_{3}$, again $d_{1}$ is chosen as the parameter that the procedure is insensitive to).  
This procedure effectively extracts from our measurements as 
much information as we are allowed to, without assuming anything about the
measuring device.
We test the \dicc{} protocol using the experimental setup described in the
previous paragraph, setting five different $\lambda$ parameters and also introducing an additional dephasing by carefully tuning the position of the calcite crystals.
The results for the $d_{2}$, $d_{3}$, $c_{3}$ parameters of the five different channels are shown in Tab.~\ref{param2}, while in Fig.~\ref{dicc} we show the correlation set, along with the optimal boundary, for channels A, C, D, and E.
To evaluate the reliability of this protocol we compare the results with the ones obtained using standard tomographic techniques (see Supplementary Material), using  again the relative difference between the union and intersection areas of the correlation sets, which is shown in Tab.~\ref{param2} as $\Delta$. In Fig.~\ref{dicc}, the gray boundaries are those estimated by the QPT on the channels, whose actions on the Bloch sphere is shown on the left side of each plot. The experimental data are shown in purple and the boundaries evaluated through the optimization algorithm are drawn in green. Fig.~\ref{dicc}c corresponds to the channel in which the additional dephasing factor is present. There is good agreement between the \dicc{}'s and the QPT's boundaries, indeed the $\Delta$ parameters are all below $3\%$.

\begin{table}[t]
\tabcolsep=2mm
\begin{tabular}{cccc}
\toprule
 & ($d_{2}$, $d_{3}$, $c_{3}$) & $\Delta$ & $\mu$\\
\colrule
\textbf{A} &  (0.735, 0.606, 0.394)  & 0.018 & 0.723\\
\textbf{B} &  (0.875, 0.789, 0.210)  & 0.009 & 0.865\\
\textbf{C} &   (0.612, 0.415, 0.585) & 0.027 & 0.833\\
\textbf{D} &   (0.823, 0.784, 0.215) & 0.009 & 0.372\\
\textbf{E} &  (0.696, 0.675, 0.325)  & 0.001 & 0.131\\
\botrule
\end{tabular}
\caption{\textbf{Device Independent Channel Characterization.} Characteristic parameters ($d_{1}$, $d_{2}$, $d_{3}$, $c_{3}$) with their statistical uncertainties obtained using the \dicc{} protocol for five different channels \textbf{A}-\textbf{E}.
In channel \textbf{E} an artificial dephasing factor was added. In the last column on the right we report the $\mu$ parameters of each device independent channel characterization, which define their equivalence class.
}
\label{param2}
\end{table}

As noted before, the \dicc{} protocol does not identify the quantum channel unambiguously \cite{dell2018}: indeed, for each minimization problem, there is an equivalence class of quantum channels, which optimizes Eq.\ref{minimize}, specified by the parameter $\mu$.
In this case, all the implemented quantum channels have $0 \leq \mu \leq 1$, as reported in Tab.\ref{param2}. This regime, which is the same that was mentioned in the previous paragraph, brings two device independent reconstructions $\mathcal{C}_{(d_{1}, d_{2}, d_{3}, \vec{c})}$ and $\mathcal{C}_{(d'_{1}, d'_{2}, d'_{3}, \vec{c}')}$ to be indistinguishable when $(d_2, d_3, c_3)= (d_2', d_3', c_3')$, as proved in \cite{dell2018}.

\textbf{\emph{Discussion.--}} In conclusion, our work provides a strong experimental insight into 
results recently derived and experimentally implemented with
superconducting qubits by Dall'Arno et al. in Refs. \cite{dell2018,ditest}. We apply those results to a photonic
implementation,  and we introduce an algorithm for the estimation of the
confidence level, in addition to that introduced in Ref.~\cite{dell2018}. Indeed, we apply two device-independent protocols derived in
Refs~\cite{dell2018,ditest} (device-independent tomography validation, \ditv{}, and device-independent channel characterization, \dicc{}) to two types of dihedrally covariant quantum channels (amplitude damping channel and dephased amplitude damping channel), implemented on a photonic platform. 
These protocols are based only on the set of input/output correlations which can be observed by an experimenter, without the need of trusting the apparatus. Through the aforementioned procedures, we were able to  validate the implemented channel reconstruction of a quantum process tomography and
to extract the maximum amount of information from the observed correlation set, in a fully device-independent way. 
This study, therefore, gives new experimental tools which can be adopted to test whether the experimental apparatus is correctly functioning, but free from the vicious cycle affecting device-dependent procedures like QPT, which require the apparatus to be trusted in the preparation of probe states and on the realization of specific projective measurements. The presented protocols do not provide a complete characterization of the implemented quantum channel (that remains undetermined up to its equivalence class), but we believe our results pave the way for future research along this direction.\\

\textbf{\textit{Acknowledgements.--}}This work was supported by the ERC-Starting Grant 3D-QUEST (3D-Quantum Integrated Optical Simulation; grant agreement number 307783): http://www.3dquest.eu. RC acknowledges the Brazilian ministries MCTIC, MEC and the CNPq. F.B.  was  supported by JSPS KAKENHI,
grant no.  17K17796. M.D. was  supported by the  Ministry of
Education and the Ministry of Manpower (Singapore).

\end{document}


\title{Supplementary Material: Experimental device-independent tests of quantum
  channels}
\author{Iris Agresti}
\affiliation{Dipartimento di Fisica - Sapienza Universit\`{a} di Roma, P.le Aldo Moro 5, I-00185 Roma, Italy}

  \author{Davide Poderini}
\affiliation{Dipartimento di Fisica - Sapienza Universit\`{a} di Roma, P.le Aldo Moro 5, I-00185 Roma, Italy}
  
  \author{Gonzalo Carvacho}
\affiliation{Dipartimento di Fisica - Sapienza Universit\`{a} di Roma, P.le Aldo Moro 5, I-00185 Roma, Italy}
  
  \author{Leopoldo Sarra}
\affiliation{Dipartimento di Fisica - Sapienza Universit\`{a} di Roma, P.le Aldo Moro 5, I-00185 Roma, Italy}

 \author{Rafael Chaves}
\affiliation{International Institute of Physics, Federal University of Rio Grande do Norte, 59078-970, P. O. Box 1613, Natal, Brazil}

\author{Francesco Buscemi}
\email{buscemi@is.nagoya-u.ac.jp}
\affiliation{Department of Mathematical Informatics, Nagoya University, Chikusa-ku, Nagoya, 464-8601, Japan}

  \author{Michele Dall'Arno}
\email{cqtmda@nus.edu.sg}
\affiliation{Centre for  Quantum Technologies, National University of  Singapore, 3 Science Drive  2, 117543, Singapore}

\author{Fabio Sciarrino}
\email{fabio.sciarrino@uniroma1.it}
\affiliation{Dipartimento di Fisica - Sapienza Universit\`{a} di Roma, P.le Aldo Moro 5, I-00185 Roma, Italy}

\date{\today}

\maketitle

In this Supplementary Material, we  apply the theoretical framework for the device-independent inference of quantum channels developed by Dall'Arno et al. in Refs.~\cite{DallArno, DBB17} to derive the mathematical boundary of the set of input/output correlations for an amplitude damping channel of efficiency $\lambda$. We also include the maximum likelihood-based method we developed to perform quantum process tomography, restricted to dihedrally covariant channels and details for the experimental channel implementation and the state generation.

\subsection{I.- Correlation set compatible with an amplitude damping channel}
\begin{figure}[hbt]
	\includegraphics[width=.33\textwidth]{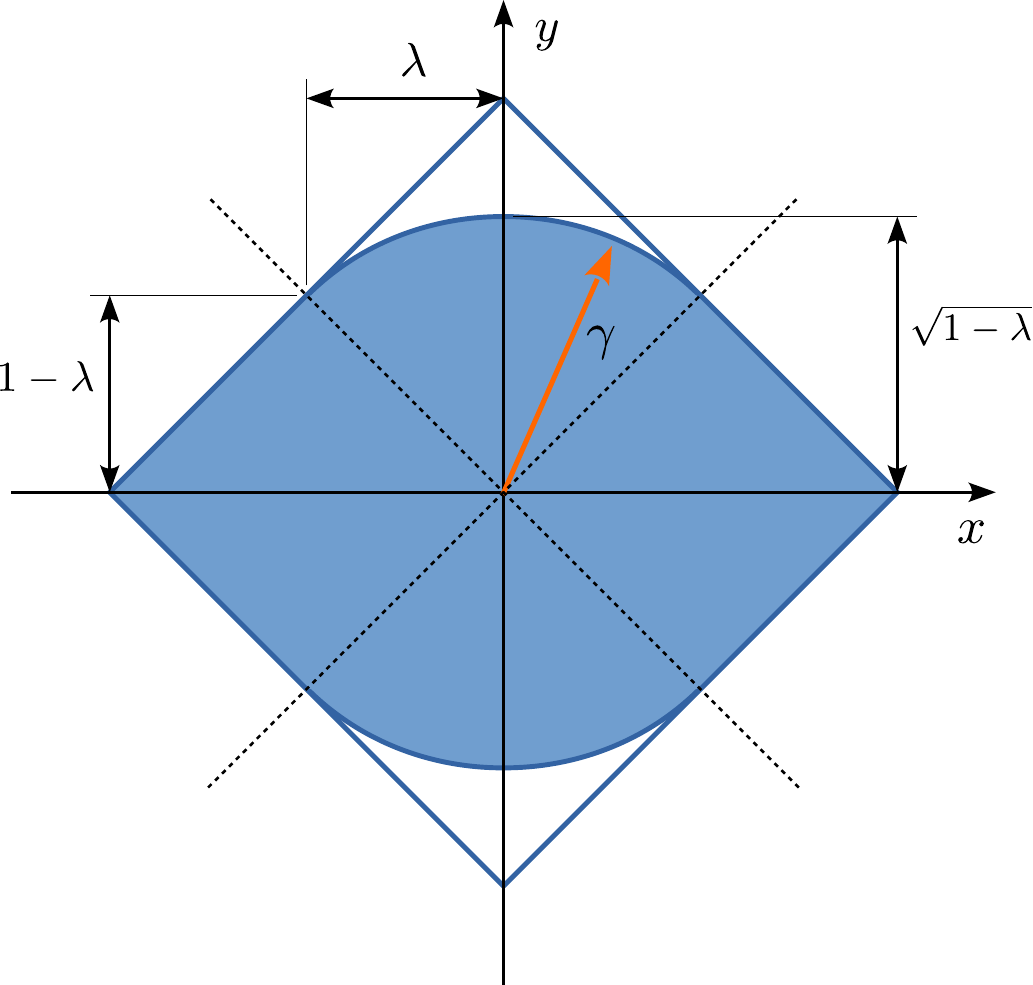}
 \caption{Cartesian representation of the set of binary input-output
   correlations compatible with the amplitude-damping channel (blue area).}
 \label{fig:cartesian}
\end{figure}

The input/output correlations that are compatible with an amplitude damping channel characterized by an efficiency $\lambda$ can be derived from \textbf{Corollary 2} of \cite{DBB17}, which states that an input-output binary probability distribution \textit{p} is compatible to an amplitude channel with efficiency $\lambda$ if and only if
\begin{align}
(\sqrt{p_{1|2}p_{2|1}}-\sqrt{p_{1|2}p_{2|1}})^2 < \lambda
  \label{eq:corollary}
\end{align}
The binary input-output probability distribution \textit{p} can be parametrized \cite{DallArno}:
\begin{align}
  \label{eq:cartesian}
  p_{j|i} = \begin{pmatrix}
    p_{1|1} & p_{2|1}\\
    p_{1|2} & p_{2|2}
  \end{pmatrix} =
             U + x X + y Y.
\end{align}
where  $|x+y| \le  1$ and  $|x-y| \le  1$, from which follows:
\begin{align*}
  x = \Tr\left[ X^T p \right], \quad y = \Tr\left[ Y^T p \right]. 
\end{align*}
Through Eq.[\ref{eq:cartesian}], inequality \ref{eq:corollary} can be expressed as:
\begin{align*}
  \frac14      \left(\sqrt{1-2y-x^2+y^2}-\sqrt{1+2y-x^2+y^2}
  \right)^{2} \leq 1- \lambda.
\end{align*}
as shown in Fig.\ref{fig:cartesian}.

Notice that the only non-trivial part of this set is for $ -\pi/4\leq \gamma  \leq  \pi/4$ and $3\pi/4\leq \gamma \leq 5\pi/4$ as all other regions form a polytope with finitely many extremal points. In this regions, the boundary can be achieved by preparing two orthogonal states
\begin{align}
    \label{eq:states}
    \begin{cases}
      \ket{\phi_0} = \sqrt{\omega}\ket{0} + \sqrt{1-\omega}\ket{1},\\
      \ket{\phi_1}       =        \sqrt{1-\omega}\ket{0}       -
      \sqrt{\omega}\ket{1},
    \end{cases}
  \end{align}
an performing a measurement described by the projector (in unnormalized form)
\begin{align}
    \label{eq:povm}
    \ket{\pi_\pm} =  \ket{0} + \frac{2  \sqrt{\lambda \omega
        (1-\omega)}}                              {2\omega-1
      \pm\sqrt{1-4\omega(1-\omega)(1-\lambda)}} \ket{1},
  \end{align}
with $\omega = 1-\tan(\gamma)/2$.

\subsection{II.- Dihedrally-Covariant Quantum Process Tomography}

Here we describe the tomographic reconstruction procedure 
restricted to return a channel within the set of $D_2$ covariant ones.
In   the    Bloch sphere
representation,  any qubit  state can be
represented as
\begin{align}
  \label{eq:bloch}
  \rho_{\vec{v}} =  \frac12 \left( \openone  + \vec{\sigma}^T
  \cdot \vec{v} \right).
\end{align}
Here, $\vec\sigma$ represents the vector of Pauli matrices and
$\| \vec{v} \| \le 1$.  Accordingly, the action of any qubit channel can be represented as
\begin{align}
  \label{eq:bloch_ch}
 \ch{X}_{A, \vec{b}} \left(  \rho_{\vec{v}} \right) :=
  \frac12 \left[  \openone +  \vec{\sigma}^T \cdot  \left( A
    \vec{v} + \vec{b} \right) \right],
\end{align}
where $A_{i,j} = \frac{1}{2}  \Tr \left[ \sigma_i \ch{X}_{A,
    \vec{b}}  \left( \sigma_j  \right) \right]$  and $b_i  =
\frac{1}{2} \Tr  \left[ \sigma_i \ch{X}_{A,  \vec{b}} \left(
  \openone \right) \right]$.  This parametrization of qubit
channels was adopted in Refs.~\cite{KR01,RSW02}.

In Ref.~\cite{DallArno}, explicit conditions for the complete positivity of the map in Eq.~\eqref{eq:bloch_ch} were derived in terms of such a parametrization.  For the sake of completeness, we report here such a derivation.  Let $\mathcal{U}$ and $\mathcal{V}$ be two qubit (anti)-unitary transformations, such that $\ch{V} \circ \ch{X}_{A, \vec{b}} \circ \ch{U}$ is a quantum channel.  One immediately has
\begin{align*}
  \ch{V}   \circ   \ch{X}_{A, \vec{b}}  \circ   \ch{U}   =
 \ch{X}_{V^T A U, V^T \vec{b}}.
\end{align*}
Here, $U,  V$ are rotations  if and  only if
$\mathcal{U}$ and $\mathcal{V}$ are unitary channels,
and the product of rotation and reflections otherwise.  If $U$ and  $V$ are some
rotations  such that $D =  V^T A U$ is  diagonal, we
set $D  = \diag(d_1,  d_2, d_3)$ and  $\vec{c} =  (c_1, c_2,
c_3) := V^T \vec{b}$. Matrices $U$ and $V$
are not uniquely specified.  By  explicit calculation, the Choi operator
$E$ of $\ch{X}_{D, \vec{c}}$ is given by
\begin{align*}
  E = \frac{1}{2}\begin{pmatrix}  1+c_3+d_3 & c_1 - i c_2 &  0 & d_1 +  d_2\\ c_1 + i c_2 &
    1-c_3-d_3  & d_1-d_2  & 0\\  0 &  d_1-d_2 &  1+c_3-d_3 &
    c_1-i c_2\\ d_1+d_2 & 0 & c_1 + i c_2 & 1-c_3+d_3
    \end{pmatrix}.
\end{align*}
A qubit channel $\ch{X}_{D,  \vec{c}}$ is $D_2$ covariant
if and  only if two entries  of $\vec{c}$ are null.
Since a rotation  matrix in $V$  and $U$  permutes the
entries of $D$ and $\vec{c}$, without loss of generality we can take $c_1 = c_2 = 0$.  
Inserting this in the Choi   operator,  
the   following  condition   for  complete positivity, which first appeared in ~\cite{DallArno},  immediately follows
\begin{align}
	\label{eq:cp}
	\begin{cases}
    	d_3 + \sqrt{\left( d_1 - d_2 \right)^2 + c_3^2} \le 1,\\
        -d_3 + \sqrt{\left( d_1 + d_2 \right)^2 + c_3^2} \le 1.
    \end{cases}
\end{align}

In general, following Ref.~\cite{hrad}, we consider an experiment in which we prepare a complete set of $M$ input states $\{\rho_i\}$ 
and perform $L$ measurements on each output state $\ch{X}_{A,\vec{b}}(\rho_i)$,
performing a complete set of POVMs $\{ \Pi_{j|i} \}$.
In particular we send the set of $\{ \ket{H}, \ket{V}, \ket{+}, \ket{-}, \ket{R}, \ket{L} \}$ as input states and the corresponding projectors as POVM.
Quantum theory gives the probability of measuring the output
$j$ given the input state $i$ as
\begin{align}
    \label{eq:quantum_probs}
    p_{j|i} = \Tr \left[ 
    \ch{X}_{A,\vec{b}} \left( \rho_i  \right)
    \Pi_{j|i}
    \right],
\end{align}
where we parametrized the channel $ \ch{X}_{A,\vec{b}}$ using
Eq.~\eqref{eq:bloch_ch}. 
In addition, we parametrize rotation matrices as
$V = \exp{\left(\sum \lambda_k m_k\right)}$ and $U = \exp{\left(\sum \lambda_k n_k\right)}$, where $\lambda_k$ are the three matrices
\begin{align*}
  \lambda_1 := \begin{pmatrix} 0 & 0 & 0 \\ 0 & 0 & -1 \\ 0 & 1 & 0 \end{pmatrix} \quad
  \lambda_2  :=\begin{pmatrix} 0 & 0 & 1 \\ 0 & 0 & 0 \\ -1 & 0 & 0  \end{pmatrix} \quad
  \lambda_3  :=\begin{pmatrix} 0 & -1 & 0 \\ 1 & 0 & 0 \\ 0 & 0 & 0  \end{pmatrix}
\end{align*}
and $m_k$, $n_k$ with $k=1,2,3$ are real parameters.

The statistics of the experiment is multinomial:
\begin{align}
\label{eq:likelihood}
    \ch{L}( \ch{X}_{A,\vec{b}}) = \frac{K!}{\prod_{i,j} n_{j|i}!} \prod_{i,j} \left(p_{j|i}\right)^{n_{j|i}},
\end{align}
where $K=ML$ is the total number of measurements and $n_{j|i} = Kf_{j|i}$ the rate of a particular outcome $j$ given the input $i$; $f_{j|i}$ are the relative frequencies. 
These frequencies approximate the theoretical probabilities given by Eq.~\eqref{eq:quantum_probs}.
We remark that this procedure assumes full knowledge of the input states and POVMs.
In order to reconstruct our channel from experimental measurements, we employ the maximum-likelihood principle.
In other words, we want to maximize the function defined in Eq.~\eqref{eq:likelihood} on a convenient space of channels, in our case those with $D_2$ symmetry.
This leaves only $10$ out of $12$ free parameters ($c_1=c_2=0$) in Eq.~\eqref{eq:likelihood} .
Ignoring the constant multinomial factor, we consider its logarithm
\begin{align}
\label{eq:loglikelihood}
  \log  \ch{L}( \ch{X}_{A,\vec{b}}) =\sum_{i,j} f_{j|i} \left(p_{j|i}\right),
\end{align}
which has the same extrema, and perform a SLSQP optimization \cite{slsqp} over its parameters, under the non-linear constraints given in Eq.~\eqref{eq:cp} in order to consider only physical maps.
To avoid finding just a local maximum, we run the numerical optimization algorithm multiple times ($1000$) with different starting points and take the parameters that give the highest maximum.

\subsection{III.- Channel implementation.}
\label{sec:channelgeneration}

\textbf{The amplitude damping channel}  was implemented by exploiting two calcite displacers that, due to their birefringence, deflect horizontally polarized light. The first calcite acts like a PBS, encoding the polarization in the path degree of freedom, since horizontally polarized photons are deflected on path ``a", while vertically polarized photons are unaffected and stay on path ``b". On path ``a" we put a half wave plate ($H_{2}$) rotated of $\pi/4$, so that the horizontal polarization becomes vertical, while on path ``b" we put another half wave plate ($H_{1}$) rotated of an angle $\beta$, which sets the efficiency of our amplitude damping channel, i.e. $\lambda$. After them, we put another calcite displacer, which again deflects the horizontal polarization towards path ``a". Finally, the two paths end into the two input modes of a beam-splitter (BS) and are recombined in a non-coherent superposition, since path ``b"'s length exceeds path ``a"'s more than the photons' coherence length. When $\beta= \pi/4$ and the input is $\ket{\pm}$=($\ket{0} \pm \ket{1})/\sqrt{2}$, path ``a" and ``b" interfere, as it is shown when measurement station projects onto $\ket{\pm}$. The results presented in this paper were achieved with 0.99 of visibility.\\\\
\textbf{The dephased amplitude damping channel} was constituted of the same optical elements as the amplitude damping channel. By changing the inclination of one calcite displacer we can unbalance the paths inside the calcite interferometer affecting in this way the phase between them and so generating a dephasing channel. Different inclinations of the calcites allows us to change continuously the visibility values of this kind of channel.
The visibility values we set were 0.95 and 0.87.\\

\subsection{IV.- State generation}
\label{sec:stategeneration1}

We generated pairs of polarization entangled photons through a spontaneous parametric down conversion process. We adopted polarization encoding, identifying the states of horizontal and vertical polarization with $\ket{0}$ and $\ket{1}$ respectively.
Among the qubit photons, we filter those that are horizontally polarized and then generate the desired state.\\
This was achieved through the sequence of a polarizing beam splitter ($PBS_{1}$), a quarter-wave plate ($G_{1}$) and a half-wave plate ($G_{2}$) (the plates were rotated respectively of $\zeta$ and $\eta$).
Particularly, for the quantum process tomography we generated the following 1-qubit states:\\ (i) $\ket{0}$ ($\zeta=0$ and $\eta=0$)\\
(ii) $\ket{1}$ ($\zeta=0$ and $\eta=\pi/4$)\\
(iii) $\ket{+}$ ($\zeta=0$ and $\eta=\pi/8$)\\
(iv) $\ket{R}=(\ket{0}+i\ket{1})/\sqrt{2}$ ($\zeta=\pi/4$ and $\eta=0$).

While for the correlation's set boundary reconstruction the input states which lie on the equator of Bloch sphere, with the form $\ket{\phi_{0}}=\sqrt{\omega} \ket{0} + \sqrt{1-\omega} \ket{1}$ and its orthogonal $\ket{\phi_{1}}$, with $\omega=(1+\tan{\gamma})^{-1}$. Those states can be generated setting $\zeta=0$ and respectively $\eta_{0}=\arccos{\sqrt{\omega}}/2$ and $\eta_{1}=\arccos{\sqrt{\omega}}/2 + \pi/4$. Then, after evolving through the channel, the state is projected onto $\ket{\pi_{0,1}}=\cos{\alpha_{0,1}}\ket{0}+ \sin{\alpha}\ket{1}$, respectively with $\alpha_{0}=2 \sqrt{\lambda \omega (1-\omega)}/(2\omega -1+ \sqrt{1-4\omega (1-\omega) (1-\lambda)})$ ($\delta =\alpha_{0}$, $\xi=  \alpha_{0}/2$) and $\alpha_{1}=\alpha_{0}+\pi/2$.